\documentclass{article}
\usepackage{spconf,amsmath,graphicx}
 \usepackage{booktabs}
 \usepackage{url}

\title{PB-LRDWWS System for the SLT 2024 Low-Resource Dysarthria Wake-Up Word Spotting Challenge}

\address{Nankai University, China\\
wangshiyao@mail.nankai.edu.cn
    }

\name{Shiyao Wang, Jiaming Zhou, Shiwan Zhao, Yong Qin$^{*}$
}
\address{Nankai University, Computer Science\\
	Tianjin, China}
\begin{document}
%
\maketitle
\renewcommand{\thefootnote}{}
\footnotetext{* Corresponding author.}
\begin{abstract}
For the SLT 2024 Low-Resource Dysarthria Wake-Up Word Spotting (LRDWWS) Challenge, we introduce the PB-LRDWWS system. This system combines a dysarthric speech content feature extractor for prototype construction with a prototype-based classification method. The feature extractor is a fine-tuned HuBERT model obtained through a three-stage fine-tuning process using cross-entropy loss. This fine-tuned HuBERT extracts features from the target dysarthric speaker's enrollment speech to build prototypes. Classification is achieved by calculating the cosine similarity between the HuBERT features of the target dysarthric speaker's evaluation speech and prototypes. Despite its simplicity, our method demonstrates effectiveness through experimental results. Our system achieves second place in the final Test-B of the LRDWWS Challenge.
\end{abstract}

\begin{keywords}
Keyword spotting, wake-up word detection, dysarthria, fine-tuning, prototype-based classification
\end{keywords}
\section{Introduction} 
\label{sec:intro}
\subsection{Related work}
Keyword spotting (KWS) systems are essential for smart devices, serving as the gateway to more advanced human-machine interactions. For individuals with dysarthria, voice-controlled devices act as invaluable assistants, empowering them to overcome mobility limitations.

Dysarthria is typically caused by conditions such as cerebral palsy, amyotrophic lateral sclerosis, or other neurological disorders. Individuals with dysarthria struggle to control their articulatory organs, resulting in speech that is often unclear and lacking fluency. Furthermore, those neurological disorders often result in paralysis, making it difficult for dysarthric individuals to interact with smart devices using keyboards or touch screens. As a result, speech has emerged as a vital means to ensure equal access to technological convenience in their daily lives.

The field of dysarthric speech recognition (DSR) has been rapidly advancing \cite{2023dsrsurvey}, with personalized DSR systems \cite{google_per_dsr,twostage_dsr,meta-learning} emerging as the primary solution. DSR systems aim to empower dysarthric speakers to interact with smart devices using simple command words \cite{uaspeech} or phrases \cite{easycall}.
However, continuous operation of DSR systems can lead to increased power consumption and a higher rate of false alarms. To mitigate these issues, integrating KWS systems as the ``initiator" of interaction between dysarthric speakers and smart devices is crucial. KWS systems can streamline the communication process while minimizing power consumption and the rate of false alarms.

In the design of KWS systems for typical speech, a primary focus is on minimizing power consumption \cite{2021kwssurvey}. These systems strive for simplicity and lightweight architecture while maintaining accuracy. Zhang et al. \cite{helloedge} introduced deep separable convolutional neural network (DS-CNN) \cite{mobilenets} from the field of computer vision to the KWS task, achieving an outstanding accuracy of 95.4\% on the widely used Speech Commands dataset \cite{speechcommand}. Additionally, Peter et al. \cite{dscnn} proposed and evaluated algorithms for training and implementing DS-CNN on embedded devices. Inspired by the two-stage object detection approach in computer vision \cite{r-fcn,fasterr-cnn}, Hou et al. \cite{twostage} presented a two-stage KWS method. Their approach combines a multi-scale depthwise temporal convolution feature extractor with a two-stage keyword detection and localization module, enabling more precise keyword detection and localization in the streaming KWS task.
\begin{figure*}[th] 
  \centering
  \includegraphics[width=\linewidth]{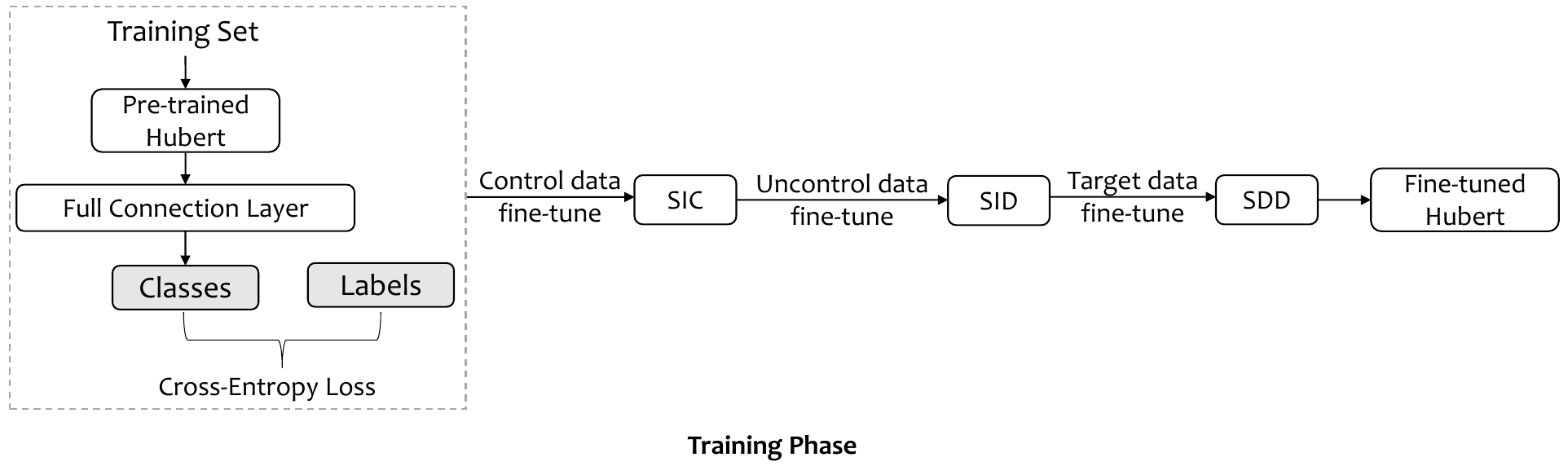}
  \caption{A three-stage fine-tuning process for building dysarthric speech content feature extractor. }
  \label{fig:f1}
  \vspace{-15pt} 
\end{figure*}
\subsection{LRDWWS Challenge}
The LRDWWS Challenge is the first initiative specifically focused on the dysarthric KWS task, aiming to detect and identify ten commonly used wake-up words. Participants in the challenge are tasked with establishing a KWS system using a limited amount of enrollment speech from each target dysarthric speaker. Then the established KWS system is used to classify the evaluation speech of the same speaker.

Taking into account the heightened challenges posed by dysarthric speech compared to typical speech, the LRDWWS Challenge has implemented task adjustments to mitigate complexity. Notably, both the enrollment speech and evaluation speech are meticulously recorded in a controlled environment, ensuring the absence of any background noise. Moreover, the dysarthric KWS task concentrates on non-streaming speech.

The evaluation metric for the dysarthric KWS task is calculated as the sum of the false acceptance rate (\textbf{FAR}) and false rejection rate (\textbf{FRR}). The formula for calculating the total score is as follows:
\begin{align}
  Score&= FRR + FAR=\frac{N_{FR}}{N_{wake}} + \frac{N_{FA}}{N_{non-wake}},
  \label{equation:eq1}
\end{align}
$N_{wake}$ represents the number of samples in the evaluation set that contain keywords, while $N_{non-wake}$ represents the number of samples without keywords. $N_{FR}$ denotes the number of samples containing keywords that are not correctly spotted by the KWS system, and $N_{FA}$ represents the number of non-keyword samples predicted as keywords. A lower score indicates better system performance.
\subsection{Our contribution} 
Given the similarity between the keywords in the dysarthric KWS task and the command words in the DSR task, we reference the prototype-based DSR (PB-DSR) work \cite{pbdsr} for addressing the dysarthric KWS task. Dysarthric speakers exhibit considerable variability in pronunciation patterns due to factors like age, etiology, severity, and speaking style. Typically, personalized DSR models are built through fine-tuning to achieve optimal DSR performance. However, the PB-DSR approach primarily relies on the prototype-based classification (PB-C) method, enabling swift adaptation of existing DSR models to unseen dysarthric speakers without extensive fine-tuning. Moreover, the PB-DSR work suggests that fine-tuning existing DSR models with the speech of the target dysarthric speaker, followed by utilizing the PB-C method, leads to optimal performance. Based on this insight, we develop a PB-LRDWWS system (depicted in Figure \ref{fig:f2}) using a three-stage fine-tuning approach (illustrated in Figure \ref{fig:f1}) and the PB-C method to effectively address the challenges in the dysarthric KWS task.

To address potential overfitting due to limited enrollment speech during fine-tuning and prevent model training failures caused by imbalanced keyword and non-keyword classes in the enrollment data, we have investigated the impact of various data augmentation techniques. These techniques include synthesizing keyword data using Text-to-Speech (TTS) systems and merging multi-stage training data into the training set. Given that the non-keyword class is not defined by a specific word or phrase, we have explored the effectiveness of different loss settings. In addition to the Connectionist Temporal Classification (CTC) loss \cite{ctcloss} used in PB-DSR, we have evaluated the performance of cross-entropy loss and the efficacy of supervised contrastive learning (SCL) \cite{scl,MADI}. Furthermore, we have compared the effects of different classification settings, including direct prediction using the fine-tuned KWS model, PB-C, and k-nearest neighbors-based classification (KNN-C).

The main contributions of this work are as follows:
\begin{itemize}
    \item We build a PB-LRDWWS system to address the challenge of the dysarthric KWS task. The PB-LRDWWS system has achieved second place in the LRDWWS Challenge. 
    \item We have compared the effects of different data augmentation and loss settings during the fine-tuning process, and also have compared the impact of different classification settings on the final classification performance. 
\end{itemize}
\section{System Overview}
\label{sec:system}
As illustrated in Figure \ref{fig:f2}, the PB-LRDWWS system comprises a dysarthric speech content feature extractor, responsible for extracting features and building prototypes, and a prototype-based classification (PB-C) method.

\subsection{Training phase: building dysarthric speech content feature extractor}
The effectiveness of the PB-LRDWWS system hinges on the representativeness of prototypes. Accurately classifying evaluation speech, based on simple metric learning (calculating cosine similarity), is only possible when the prototypes are distinguishable and representative. Since prototypes are derived by averaging sample features within each class, a robust feature extractor is crucial. To achieve this, we employ a three-stage fine-tuning method, illustrated in Figure \ref{fig:f1}, to obtain the desired dysarthric speech content feature extractor. The three-stage fine-tuning process is as follows:
\begin{itemize} 
    \item Stage 1: The pre-trained HuBERT \cite{HuBERT} model is fine-tuned using non-dysarthric speech (\textbf{Control data}) to build a speaker-independent control (\textbf{SIC}) model. 
    
    \item Stage 2: The \textbf{SIC} model is further fine-tuned using speech data from multiple dysarthric speakers (\textbf{Uncontrol data}) to build a speaker-independent dysarthria (\textbf{SID}) model. Importantly, the \textbf{Uncontrol data} does not include speech from the target dysarthric speaker.
    
    \item Stage 3: The \textbf{SID} model is further fine-tuned using the target dysarthric speaker's speech data (\textbf{Target data}) to build a speaker-dependent dysarthria (\textbf{SDD}) model. The fine-tuned HuBERT model from the \textbf{SDD} model then serves as the dysarthric speech content feature extractor.
\end{itemize}

\subsection{Inference phase: PB-LRDWWS} 
\begin{figure}[t] 
  \centering
  \includegraphics[width=\linewidth]{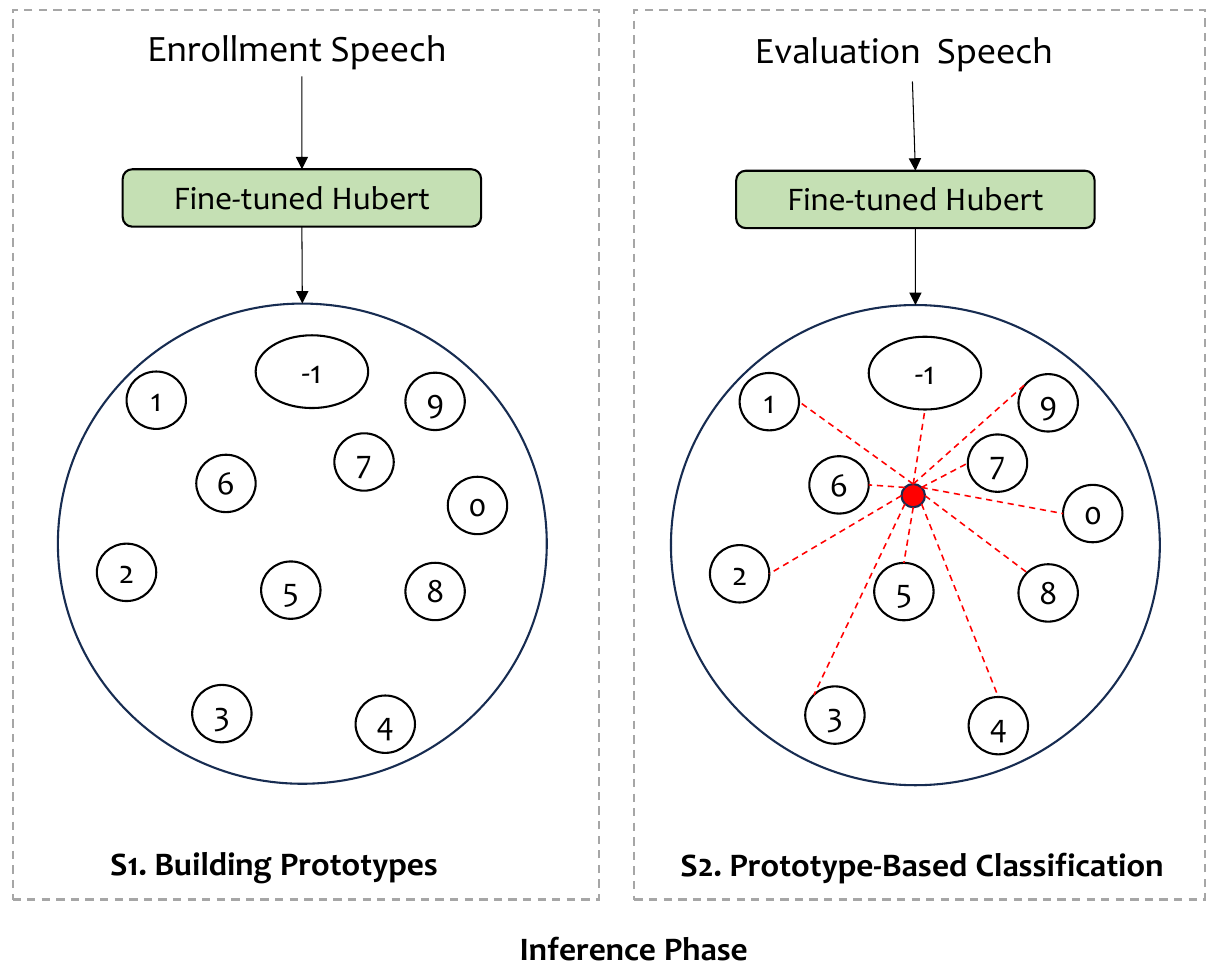}
  \caption{PB-LRDWWS.}
  \label{fig:f2}
  \vspace{-20pt}
\end{figure}

The utilization of the PB-LRDWWS system, as illustrated in Figure \ref{fig:f2}, involves two steps.

The first step involves building prototypes. Using the dysarthric speech content feature extractor, features are extracted from the enrollment speech of the target dysarthric speaker. Sample features are then averaged for each class to build prototypes.

The second step involves prototype-based classification. Using the dysarthric speech content feature extractor, the test feature is extracted from the target dysarthric speaker's evaluation speech. The cosine similarity between the test feature and each of the eleven prototypes is then calculated. The class represented by the prototype with the highest similarity is selected as the classification result.
\section{Implementation Details}
\label{sec:details}
\subsection{Dataset}
The LRDWWS Challenge is structured in three stages, each involving the release of different datasets. The datasets released for each stage are outlined below, with the important note that the speakers in all datasets are not overlapping.
\begin{itemize}
    \item Stage 1 involves the release of two datasets: LRDWWS\_train.zip and LRDWWS\_dev.zip. LRDWWS\_tr-ain.zip contains speech from control speakers (\textbf{DC}) and speech from multiple dysarthric speakers (\textbf{DUC}), while LRDWWS\_dev.zip includes speech from four dysarthric speakers (\textbf{Dev4}).
    
    \item Stage 2 involves the release of LRDWWS\_test.zip, containing speech from ten dysarthric speakers (\textbf{Test-A}).
    
    \item Stage 3 involves the release of LRDWWS\_test\_b.zip, containing speech from ten dysarthric speakers (\textbf{Test-B}). Importantly, the evaluation speech data in \textbf{Test-B} is unlabeled.
\end{itemize}
\renewcommand{\thefootnote}{\arabic{footnote}} 
During fine-tuning, the splits of \textbf{DC} and \textbf{DUC} data remain consistent with the \textbf{Baseline}\footnote{\url{https://github.com/greeeenmouth/LRDWWS}}. For \textbf{Dev4}, \textbf{Test-A}, and \textbf{Test-B}, the enrollment speech of each target dysarthric speaker is used for training and validation, while the evaluation speech serves as the test set.

\begin{table*}[th]
    \caption{The performance of \textbf{SIC} models and \textbf{SID} models on \textbf{Dev4}. ``DC\_train'' indicates the training set of \textbf{DC}. ``DUC\_train'' indicates the training set of \textbf{DUC}. This table is organized in descending order based on the ``\textbf{Dev4 Score}''.}
    \label{tab:t1}
    \centering
    \begin{tabular}{c | l | c}
        \toprule
        \textbf{Model ID} & \textbf{Fine-tuning Setting} &	\textbf{Dev4 Score}\\   
        \midrule
        \textbf{SIC-1} & DC\_train + \textbf{CTC} & Model Crash \\
        \textbf{SIC-2} & DC\_train + \textbf{TTS} + \textbf{CTC} & 0.49389 \\
        \textbf{SIC-3} & DC\_train + \textbf{TTS} + \textbf{CTC} + \textbf{AddSCL} & 0.39375 \\
        \textbf{SIC-4} & DC\_train + \textbf{TTS} + \textbf{CE} & 0.31347 \\
        \textbf{SIC-5} & DC\_train + \textbf{CE} & 0.27611 \\
        \textbf{SIC-6} & DC\_train + \textbf{CE} + \textbf{AddSCL} & 0.23846\\
        \textbf{SID-1} & Fine-tune \textbf{SIC-6} + DUC\_train + \textbf{CE} + \textbf{AddSCL} & 0.15653 \\
        \textbf{SID-2} & Fine-tune \textbf{SIC-5} + DUC\_train + \textbf{CE} & 0.15625 \\
        \textbf{SID-3} & Fine-tune \textbf{SIC-3} + \textbf{Merge\_train} + \textbf{CTC} + \textbf{AddSCL} & 0.12507 \\  
         \bottomrule
    \end{tabular}
    \vspace{-15pt} 
\end{table*}
\subsection{Experimental setup}
\label{sec:exp_setup}

\noindent\textbf{Basic training settings}

We fine-tune a pre-trained HuBERT model\footnote{\url{https://huggingface.co/TencentGameMate/chinese-hubert-base}} under the Fairseq open-source framework\footnote{\url{https://github.com/facebookresearch/fairseq/blob/main/examples/hubert/README.md}} to create a dysarthric KWS model. For fine-tuning, we modify the ``base\_10h.yaml" configuration file\footnote{\url{https://github.com/facebookresearch/fairseq/blob/main/examples/hubert/config/finetune/base_10h.yaml}}, setting the learning rate to $10^{-5}$ and the warmup steps to 32,000. Training is stopped when the training loss does not decrease for 10 consecutive epochs.

\noindent\textbf{Data augmentation settings}
\begin{itemize}
    \item \textbf{TTS}: We train a multi-speaker TTS system using FastSpeech2 \cite{fastspeech2} with speech data from 174 speakers in the AISHELL-3 dataset \cite{aishell3}. This TTS system is then used to synthesize all Chinese keyword speech data, which is subsequently added to the training set.
    \item \textbf{Merge\_train}: We create a merged training set, denoted as ``\textbf{Merge\_train}", by combining the \textbf{TTS} data with the training data from both \textbf{DC} and \textbf{DUC}.
\end{itemize}

\noindent\textbf{Loss settings}
\begin{itemize}
    \item \textbf{CTC}: We train the dysarthric KWS model using the CTC loss \cite{ctcloss}. The model's dictionary includes special tokens like $<sos>, <eos>, <pad>$, and blank, along with 10 keyword IDs (0-9) and a non-keyword ID (-1).
    \item \textbf{CE}: We train the dysarthric KWS model using cross-entropy loss for classification. The model distinguishes between 10 keyword classes and one non-keyword class. The cross-entropy loss, denoted as $L_{CE}$, is formulated as follows:
        \begin{align}
          L_{CE} &=  \sum_{i \in I } (y_ilog\hat{y_i} +(1-y_i)log(1-\hat{y_i})),
          \label{equation:eq2}
        \end{align}
    where $i \in I = \{1, . . . , N\}$ represents the index of a speech, $y_i$ denotes the true label of speech $i$, and $\hat{y_i}$ represents the model's predicted label for speech $i$.
    \item \textbf{AddSCL}: We combine CTC loss $L_{CTC}$ and cross-entropy loss $L_{CE}$ with SCL loss $L_{SCL}$ to enhance training as follows:
        \begin{align}
          L_{Total_1}&= L_{CTC} + L_{SCL}, \\
          L_{Total_2}&= L_{CE} + L_{SCL}.
          \label{equation:eq3}
        \end{align}
    In SCL, within a training batch, samples with identical labels are considered positive pairs, while samples with distinct labels are considered negative pairs. The SCL loss is formulated as follows:
        \begin{align}
          L_{SCL} &=  \sum_{i \in I } \frac{-1}{|P(i)|} \sum_{p \in P(i) } log \frac{exp(x_i · x_p)/\tau}{\sum_{a \in A(i)  }exp(x_i · x_p)/\tau},
          \label{equation:eq5}
        \end{align}
    where $i \in I = \{1, . . . , N\}$ represents the index of a speech, $A(i)$ denotes the set of all indices except $i$, and $x_{i}$ represents the feature extracted by HuBERT for speech $i$. $P(i)$ denotes the set of indices for all positive samples of sample $i$. Features representing the same word as sample $i$ are considered positive samples. $\tau$ is the temperature hyperparameter, which we set to 0.07.
\end{itemize}

\begin{table*}[th]
    \caption{The performance of dysarthric KWS systems on \textbf{Test-A} and \textbf{Test-B}.}
    \label{tab:t2}
    \centering
    \begin{tabular}{l | l | c c c | c c c}
        \toprule
        \textbf{Fine-tuning} & \textbf{Classification} & & \textbf{Test-A} & & & \textbf{Test-B} &\\  
        \textbf{Setting} & \textbf{Setting} & \textbf{Score} & \textbf{FAR} & \textbf{FRR} & \textbf{Score} & \textbf{FAR} & \textbf{FRR} \\
        \midrule
        \textbf{Baseline} & --- & 0.3112 & 0.0387 & 0.2725 & 0.130306 & 0.028639 & 0.101667 \\
        \hline
        Fine-tune \textbf{SID-3} + \textbf{CTC} + \textbf{AddSCL} & \textbf{PB-C} & 0.0643 & 0.0043 & 0.0600 & 0.012249 & 0.004749 & 0.007500\\
            & \textbf{KNN-C} & 0.0626 & 0.0051 & 0.0575 & 0.011812 & 0.004312 & 0.007500\\
            & \textbf{Model prediction} & 0.0645	& 0.0045 & 0.0600 & 0.014312 & 0.004312 & 0.010000 \\
        \hline
        Fine-tune \textbf{SID-2} + \textbf{CE} & \textbf{PB-C} & \textbf{0.0624} & 0.0049 & 0.0575 & \textbf{0.009801} & 0.004801 & 0.005000 \\
            & \textbf{KNN-C} & 0.0696 & 0.0046 & 0.0650 & 0.017197 & 0.004697 & 0.012500 \\
            & \textbf{Model prediction} & 0.0648	& 0.0048 & 0.0600 & 0.014851 & 0.004851 & 0.010000\\
         \bottomrule
    \end{tabular}
    \vspace{-15pt} 
\end{table*}

\noindent\textbf{Classification settings}
\begin{itemize}
    \item \textbf{PB-C}: We build prototypes by averaging the features of enrollment speech within each class. Referring to PB-DSR, we have conducted a comparison between building prototypes using full Hubert features and building prototypes using only the first frame of Hubert features. Our findings indicate that the latter setting consistently yields superior results. Therefore, we use only the first frame of Hubert features for prototype construction in our system. Classification is performed by calculating the cosine similarity between the test feature and each prototype. The label of the prototype with the highest similarity is assigned as the classification result.
    \item \textbf{KNN-C}: We perform classification by calculating the cosine similarity between the test feature and the feature of each enrollment speech from the target dysarthric speaker. The label of the enrollment speech with the highest similarity is assigned as the classification result.
    \item \textbf{Model prediction}: We use the corresponding \textbf{SDD} model to predict the class of the target dysarthric speaker's evaluation speech.
\end{itemize}

\section{Evaluations and results}
\label{sec:eval}

\subsection{Building speaker-independent dysarthria models with better generalization ability}
In the first stage of the LRDWWS Challenge, we combine the data augmentation and loss settings described in Section \ref{sec:exp_setup} to build speaker-independent dysarthria (\textbf{SID}) models with improved generalization performance. We evaluate the performance of both speaker-independent control (\textbf{SIC}) and \textbf{SID} models on the \textbf{Dev4} dataset, with the specific results summarized in Table \ref{tab:t1}.

Table \ref{tab:t1} reveals several key findings. Firstly, the class imbalance between keywords and non-keywords in the training data can lead to model instability when using CTC loss (model \textbf{SIC-1}), while training proceeds normally under cross-entropy loss (model \textbf{SIC-5}). Secondly, adding TTS data to the training set mitigates the impact of class imbalance on CTC loss training (model \textbf{SIC-2}) but negatively affects performance under cross-entropy loss (model \textbf{SIC-4}). This difference in effect can be attributed to the underlying task differences between CTC loss and cross-entropy loss. CTC loss, used for the recognition task, benefits from diverse samples, while cross-entropy loss, used for the classification task, is sensitive to large differences within the same category. Finally, incorporating SCL loss improves the performance of the \textbf{SIC} model under both CTC loss (model \textbf{SIC-2} vs. model \textbf{SIC-3}) and cross-entropy loss (model \textbf{SIC-5} vs. model \textbf{SIC-6}). This improvement is attributed to SCL's ability to enhance the representativeness and distinguishability of features by bringing features of the same category closer and pushing features of different categories further apart during model training, thereby improving performance.

We build \textbf{SID} models based on \textbf{SIC} models \textbf{SIC-5} and \textbf{SIC-6}, which demonstrate superior performance among the \textbf{SIC} models, as well as model \textbf{SIC-3}, which achieves the best performance under the CTC loss setting. The fine-tuning settings for the \textbf{SID} models are kept consistent with their corresponding \textbf{SIC} models. After fine-tuning with \textbf{Uncontrol data}, we observe that even without including the speakers from \textbf{Dev4} in the training data, the models effectively have learned common features of dysarthric speech, leading to a significant performance improvement on \textbf{Dev4} (model \textbf{SIC-6} vs. model \textbf{SID-1}, model \textbf{SIC-5} vs. model \textbf{SID-2}, and model \textbf{SIC-3} vs. model \textbf{SID-3}).

\subsection{Building speaker-dependent dysarthria models and comparing the effects of different classification settings}
Due to the presence of more dysarthric speakers in \textbf{Test-A} compared to \textbf{Dev4}, we fine-tuned \textbf{SID} models \textbf{SID-2} and \textbf{SID-3} using speech data from \textbf{Test-A}. The evaluation results returned by the LRDWWS Challenge evaluation system are presented in Table \ref{tab:t2}.

Table \ref{tab:t2} compares the performance of three classification settings. Our results demonstrate a significant improvement over the \textbf{Baseline}. Under the CTC loss setting with SCL loss, the \textbf{KNN-C} method achieves the best performance, while \textbf{PB-C} and \textbf{Model prediction} exhibit comparable results. The superiority of \textbf{KNN-C} can be attributed to SCL's direct influence on sample features, bringing features from the same class closer together and pushing features from different classes further apart. In contrast, under the cross-entropy loss setting, the \textbf{PB-C} method proves optimal, while \textbf{KNN-C} performs the worst. One possible explanation is that cross-entropy loss, being a classification loss, does not directly influence sample features. Consequently, using prototypes derived from averaging sample features for classification leads to more stable performance.

\subsection{Final result}
We apply all the settings tested on \textbf{Test-A} to \textbf{Test-B} and submit the results to the LRDWWS Challenge evaluation system. The evaluation results are presented in Table \ref{tab:t2}. We can observe that the results of the six submissions for \textbf{Test-B} are consistent with the performance trends observed on \textbf{Test-A}. The best performance is still achieved through fine-tuning using cross-entropy loss and classifying using the \textbf{PB-C} method. The comparison results of the different classification settings under each fine-tuning setting are consistent with the conclusions drawn from \textbf{Test-A}.

\section{Conclusion}
We propose the PB-LRDWWS system for the dysarthric KWS task introduced in the SLT 2024 LRDWWS Challenge. The PB-LRDWWS system comprises a dysarthric speech content feature extractor for building prototypes and a prototype-based classification method. The PB-LRDWWS system has achieved a score of \textbf{0.009801} on the final \textbf{Test-B}, significantly outperforming the \textbf{Baseline} and securing second place in the LRDWWS Challenge.



\section{ACKNOWLEDGMENTS}
\label{sec:ack}
This work has been supported in part by  NSF China (Grant No.62271270).

\bibliographystyle{IEEEbib}
\bibliography{refs}

\end{document}